\documentstyle[12pt,epsf]{article}

\newcommand{\ct}{\cite}
\newcommand{\lb}{\label}

\newcommand{\bc}{\begin{center}}
\newcommand{\ec}{\end{center}}
\newcommand{\bd}{\begin{displaymath}}
\newcommand{\ed}{\end{displaymath}}
\newcommand{\be}{\begin{equation}}
\newcommand{\ee}{\end{equation}}
\newcommand{\ba}{\begin{array}}
\newcommand{\ea}{\end{array}}
\newcommand{\bt}{\begin{tabular}}
\newcommand{\et}{\end{tabular}}

\newcommand{\ov}{\overline}
\newcommand{\bp}{\begin{picture}}
\newcommand{\ep}{\end{picture}}
\newcommand{\bfi}{\begin{figure}}
\newcommand{\efi}{\end{figure}}

\begin{document}

\title {\Large \bf The Fundamental-Weak Scale Hierarchy in the Standard Model}

\author{{\large C.D.Froggatt}${}^{1}$,{\large ~ L.V.Laperashvili}${}^{2}$,
{\large ~ H.B.Nielsen}${}^{3}$\\[10mm] \itshape{${}^{1}$ Department of
Physics and Astronomy,}\\[3mm] \itshape{Glasgow University,
Glasgow, Scotland.}\\[3mm] \itshape{${}^{2}$ Institute of Theoretical and Experimental Physics,
 Moscow, Russia.}\\[3mm] \itshape{${}^{3}$ The Niels Bohr
Institute, Copenhagen, Denmark. }}

\date{}

\maketitle

\thispagestyle{empty}

\begin{abstract}

The multiple point principle, according to which several vacuum
states with the same energy density exist, is put forward as a
fine-tuning mechanism predicting the ratio between the fundamental
and electroweak scales in the Standard Model (SM). It is shown
that this ratio is exponentially huge: $\sim e^{40}$. Using
renormalisation group equations for the SM, we obtain the
effective potential in the 2-loop approximation and investigate
the existence of its postulated second minimum at the fundamental
scale. The investigation of the evolution of the top quark Yukawa
coupling constant in the 2-loop approximation shows that, with
initial values of the top Yukawa coupling in the interval
$h(M_t)=0.95\pm 0.03$ (here $M_t$ is the top quark pole mass), a
second minimum of the SM effective potential can exist in the
region $\phi_{min2}\approx 10^{16}-10^{22}$ GeV. A prediction is
made of the existence of a new bound state of 6 top quarks and 6
anti-top quarks, formed due to Higgs boson exchanges between pairs
of quarks/anti-quarks. This bound state is supposed to condense in
a new phase of the SM vacuum. This gives rise to the possibility
of having a phase transition between vacua with and without such a
condensate. The existence of three vacuum states (new, electroweak
and fundamental) solves the hierarchy problem in the SM.

\end{abstract}

\vspace{1cm} \footnoterule{\noindent${}^{1}$ E-mail:
c.froggatt@physics.gla.ac.uk\\ ${}^{2}$ E-mail:
laper@heron.itep.ru\\ ${}^{3}$ E-mail: hbech@nbi.dk}

\newpage

\pagenumbering{arabic}

\section{Introduction: cosmological constant and multiple-point principle}

One of the main goals of physics today is to find the fundamental theory
beyond the Standard Model (SM). The vast majority of the available
experimental information is already explained by the SM.
Until now, there is no evidence for the existence of any particles
or bound states composed of new particles other than those of the
SM. All accelerator physics is in agreement with the SM, except
for neutrino oscillations. Presently only this neutrino physics, together
with astrophysics and cosmology, gives us any
phenomenological evidence for going beyond the SM.

In first approximation one might ignore these indications of new
physics and consider the possibility that the SM essentially represents
physics well up to the Planck scale. In the present paper,
developing the ideas of Ref.~\ct{1},
we suggest a scenario, using only the pure SM, in which an
exponentially huge ratio between the fundamental (Planck) and
electroweak scales results:
$$
     \frac{\mu_{fundamental}}{\mu_{electroweak}} \sim e^{40}.
$$
This exponentially huge scale ratio occurs due to the required
degeneracy of the three vacuum states discussed in Refs.[2,3].

In such a scenario it is reasonable to assume the existence of a
simple and elegant postulate which helps us to explain the SM
parameters: couplings, masses and mixing angles. In our model such
a postulate is based on a phenomenologically required result in
cosmology \ct{4}: the cosmological constant is zero, or
approximately zero, meaning that the vacuum energy density is
very small. {\it A priori} it is quite possible for a quantum field
theory to have several minima of its effective potential as a
function of its scalar fields. Postulating zero cosmological
constant, we are confronted with a question: is the energy density, or
cosmological constant, equal to zero (or approximately zero) for
all possible vacua or it is zero only for that vacuum in which we
live?

This assumption would not be more complicated if we postulate that
all the vacua which might exist in Nature, as minima of the
effective potential, should have approximately zero cosmological
constant. This postulate corresponds to what we call the Multiple
Point Principle (MPP) [5,6].

MPP postulates: {\it there are many vacua with the same energy
density or cosmological constant, and all cosmological constants
are zero or approximately zero.}

There are circa 20 parameters in the SM characterizing the
couplings and masses of the fundamental particles, whose values
can only be understood in speculative models extending the SM. In
Ref.~\ct{7} it was shown that the Family Replicated Gauge Group
(FRGG) model, suggested in Refs.~[8,9] as an extension of the SM
(see also the reviews [10,11]), fits the SM fermion masses and
mixing angles and describes all neutrino experimental data order
of magnitudewise using only 5 free parameters -- five vacuum
expectation values of the Higgs fields which break the FRGG
symmetry to the SM. This approach based on the FRGG--model was
previously called Anti--Grand Unified Theory (AGUT) and developed
as a realistic alternative to SUSY Grand Unified Theories (GUTs)
~[12-16]. In Refs.~[17,18] the MPP was applied to the
investigation of phase transitions in regularized gauge theories.
It was shown in \ct{19} that MPP forbids the existence of a fourth
generation. A tiny order of magnitude of the cosmological constant
was explained in a model involving supersymmetry breaking in N=1
supergravity and MPP \ct{20}. An investigation of the hierarchy
problem in the SM extended by MPP and two Higgs doublets is in
progress \ct{21}.

In the present paper we use MPP with the aim of solving the
hierarchy problem, in the sense that we give a crude prediction
for the fundamental to electroweak scale ratio in the pure SM. It
is necessary to emphasize that this result essentially depends on
predicting the value of the top quark Yukawa coupling constant
consistent with experiment at the electroweak scale. This
prediction entails the existence of a new bound state of 6 top
quarks and 6 anti-top quarks, which condenses in a new phase of
the SM vacuum ~[2,3,7]. In addition we require the existence of a
third SM vacuum, with a Higgs field expectation value of the order
of the fundamental scale, which we discuss first in terms of the
renormalization group improved potential.

\section{The renormalization group equation for the effective
potential}

\subsection{The Callan-Symanzik equation}

In the theory of a single scalar field interacting with a gauge
field, the effective potential $V_{eff}(\phi_c)$ is a function of
the classical field $\phi_c$ given by
\be
       V_{eff} = - \sum_0^\infty
       \frac{1}{n!}\Gamma^{(n)}(0)\phi_c^n,     \lb{1}
\ee where $\Gamma^{(n)}(0)$ is the one-particle irreducible (1PI)
n-point Green's function calculated at zero external momenta.

The renormalization group equation (RGE) for the effective
potential means that the potential cannot depend on a change in
the arbitrary renormalization scale parameter M:
\be
         \frac{dV_{eff}}{dM} = 0.           \lb{2}
\ee
The effects of changing it are absorbed into changes in the
coupling constants, masses and fields, giving so-called running
quantities.

Considering the renormalization group (RG) improvement of the
effective potential [22,23] and choosing the evolution variable as
\be
          t = \log(\mu/M) = \log(\phi/M),    \lb{3}
\ee where $\mu$ is the energy scale, we have the Callan-Symanzik
[24,25] RGE for the full $V_{eff}(\phi_c)$ with $\phi\equiv
\phi_c$ : \be
     (M\frac{\partial}{\partial M} + \beta_{m^2}\frac{\partial}
     {\partial m^2} + \beta_{\lambda}
     \frac{\partial}{\partial \lambda} + \beta_g \frac{\partial}{\partial g}
      + \gamma\, \phi\frac{\partial}{\partial \phi})
     V_{eff}(\phi) = 0,                            \lb{4}
\ee
where M is a renormalization mass scale parameter, $\beta_{m^2}$,
$\beta_{\lambda}$, $\beta_g$ are the RG beta functions for the
scalar mass squared $m^2$, the scalar field self-interaction $\lambda$
and the gauge couplings respectively. Also  $\gamma$ is the anomalous
dimension, and the set of gauge coupling constants for the SM are:
$g_i = (g', g, g_3)$ for the $U(1)_Y$, $SU(2)_L$ and $SU(3)_c$ groups.
Here the couplings depend on the renormalization scale M:
$\lambda = \lambda(M)$, $m^2 = m^2(M)$ and $g_i = g_i(M)$. In the
following we shall also introduce the top quark Yukawa coupling
$h\stackrel{def}{=}g_t$ and neglect the Yukawa couplings of all the
lighter fermions.

It is convenient to introduce a more compact notation for the
parameters of the theory.
Define:
\be
          \lambda_p = (m^2,\, \lambda,\, g),    \lb{5}
\ee
so that the RGE  can be abbreviated as
\be
   (M\frac{\partial}{\partial M} + \beta_p\frac{\partial}{\partial \lambda_p} +
     \gamma \, \phi \frac{\partial}{\partial \phi})V_{eff}
     = 0.                                                     \lb{6}
\ee
The general solution of the above-mentioned RGE has the
following form \ct{22}:
\be
  V_{eff} = - \frac{m^2(\phi)}{2}[G(t)\phi]^2
            + \frac{\lambda(\phi)}{8}[G(t)\phi]^4 + C,  \lb{6a}
\ee
where
\be
            G(t) = \exp(-\int_0^t \gamma(t')dt').           \lb{6b}
\ee
We shall also use the notation $\lambda(t) = \lambda(\phi)$,
$m^2(t) = m^2(\phi)$, $g_i^2(t) = g_i^2(\phi)$, which should not lead
to any misunderstanding.
In the loop expansion $V_{eff}$ is given by
\be
        V_{eff} = V^{(0)} + \sum_{n=1}^{\infty} V^{(n)},     \lb{7}
\ee where $V^{(0)}$ is the tree-level potential of the SM.
Similarly the RG $\beta$-functions have the expansion:
\be
     \beta_p = \sum_{n=1}^{\infty} \beta_p^{(n)},\quad \gamma =
     \sum_{n=1}^{\infty}\gamma^{(n)},                            \lb{8}
\ee
where $X^{(n)}$ is the n-loop contribution to X.

\newpage

Finally, following Sher's method \ct{23}, we have the
following equations for the arbitrary loop approximation:
\be
    M\frac{\partial}{\partial M}V^{(1)} + D_1 V^{(0)} = 0
                                                     \lb{9}
\ee
- for the one-loop approximation,
\be
    M\frac{\partial}{\partial M}V^{(2)} + D_2 V^{(0)} + D_1 V^{(1)} = 0
                                                     \lb{10}
\ee - for the two-loop approximation, {\bc \large \bf

.

.

. \ec}
 \be
    M\frac{\partial}{\partial M}V^{(n)} + D_n V^{(0)} + D_{n-1} V^{(1)}
     + ... + D_1V^{(n-1)} = 0
                                                     \lb{11}
\ee
- for the n-loop approximation.
Here the differential operator $D_k$ is defined by
\be
D_k = \beta_p^{(k)}\frac{\partial}{\partial \lambda_p} -
\gamma^{(k)}\phi \frac{\partial}{\partial \phi},\quad k=1,2,3,...
                       \lb{12}
\ee
So we have a recursion formula for the calculation of
the n-loop contribution to $V_{eff}$, using the tree-level
potential and RG functions.

\subsection{The tree-level Higgs potential}

The Higgs mechanism is the simplest mechanism leading to the
spontaneous symmetry breaking of a gauge theory. In the SM the
breaking
\be
       SU(2)_L\times U(1)_Y \to U(1)_{em},          \lb{13}
\ee achieved by the Higgs mechanism, gives masses to the gauge
bosons $W^{\pm}$, $Z$, the Higgs boson and the fermions.

With one Higgs doublet of $SU(2)_L$, we have the following
tree--level Higgs potential:
\be
        V^{(0)} = - m^2 \Phi^{+}\Phi + \frac{\lambda}{2}
        (\Phi^{+}\Phi )^2.                  \lb{14}
\ee The vacuum expectation value of $\Phi$ is:
\be
              <\Phi> = \frac{1}{\sqrt 2}\left(
             \ba{c}
             0\\
             {\it v}
             \ea
             \right),               \lb{15}
\ee where
\be
 v = \sqrt{\frac{2 m^2}{\lambda}}\approx 246\,\,{\mbox{GeV}}.
                                      \lb{16}
\ee
Introducing a four-component real Higgs field $\phi$ normalised
such that
\be
      \Phi^{+}\Phi = \frac{1}{2}\phi^2,    \lb{17}
\ee where
\be
       \phi^2 = \sum_{i=1}^4 \phi_i^2,     \lb{18}
\ee
we have the following tree-level potential:
\be
     V^{(0)} = - \frac{1}{2} m^2 \phi^2 + \frac{1}{8} \lambda
     \phi^4.              \lb{19}
\ee
As is well-known, the masses of the gauge bosons $W$ and $Z$,
a fermion with flavor $f$ and the physical Higgs boson $H$ are
expressed in terms of the VEV parameter $v$:
\be
          M_W^2 = \frac{1}{4} g^2 v^2,         \lb{20}
\ee
\be
          M_Z^2 = \frac{1}{4} (g^2 + g'^2) v^2,         \lb{21}
\ee
\be
          m_f = \frac{1}{\sqrt 2} h_f v,         \lb{22}
\ee
\be
          M_H^2 = \lambda v^2,         \lb{23}
\ee
where $h_f$ is the Yukawa coupling for the fermion with flavor $f$.

\subsection{The two-loop SM effective potential}

In the SM we use RGEs with $\beta$-functions:
\be
             \beta_{\lambda_p} = \frac{d\lambda_p}{dt}
                                     \lb{24}
\ee
given by Ref.~\ct{26} in the one-loop
and two-loop approximations (see the Appendix).

Using these $\beta$-functions, it is easy to calculate the
one--loop effective potential \ct{23}:
$$
 V_{eff}(1-loop) = - \frac{1}{2} m^2 \phi^2 +
\frac{\lambda}{8}\phi^4 + \frac{1}{2} a \phi^4
\log(\frac{\phi^2}{M^2}) +
$$
\be
\frac{1}{64\pi^2}[(m^2 -
\frac{3}{2}\lambda \phi^2)^2\log(\frac{ - m^2 +
 \frac{3}{2}\lambda \phi^2}{M^2}) + ( m^2 - \frac{1}{2}\lambda
\phi^2)^2\log(\frac{- m^2 + \frac{1}{2}\lambda \phi^2}{M^2})] + C,
                                  \lb{42}
\ee
where C is a constant,
\be
   a = \frac{1}{32\pi^2}(\frac{3}{8}g'^4 + \frac{3}{4}g'^2g^2 +
\frac{9}{8}g^4 - 6h^4)
                          \lb{43}
\ee
and the couplings are evaluated at the renormalization scale M.
Here radiative corrections due to the scalar mass term are
taken into account.

Following the same procedure \ct{23} of imposing a loop expansion on
the RGE of $V_{eff}$ and using all the RGEs \ct{26}, we have calculated
the 2--loop effective potential in the limit:
\be
      \phi^2 >> v^2, \quad\quad \phi^2 >> m^2.     \lb{44}
\ee
In general, $V_{eff}$ is given by the following series:
\be
         V_{eff}\approx V^{(0)} + V^{(1)} + V^{(2)} + ...
                                       \lb{45a}
\ee
Neglecting the radiative corrections due to the scalar mass
term, we obtain the following expression for $V^{(2)}$:
\be
    M\frac{\partial}{\partial M}V^{(2)} = - A^{(2)} \phi^4
    - B^{(2)} \phi^4 \log(\frac{\phi^2}{M^2}),  \lb{45}
\ee
where
\be
     A^{(2)} = \frac{1}{8}\gamma^{(1)}(\beta_{\lambda}^{(1)} +
     4\lambda \,\gamma^{(1)}) +
      \frac{1}{2}\lambda \, \gamma^{(2)}
      + \frac{1}{8}\beta_{\lambda}^{(2)},            \lb{46}
\ee
$$
     B^{(2)} = \frac{1}{4}\gamma^{(1)}(\beta_{\lambda}^{(1)} +
     4\lambda \,\gamma^{(1)}) + \frac{3}{32\pi^2}\lambda\,\beta_{\lambda}^{(1)}
 +$$
\be
      \frac{3}{256\pi^2}\beta_{g'}^{(1)}(g'^3 + g'g^2)
     + \frac{3}{256\pi^2}\beta_{g}^{(1)}(3g^3 + g'^2g) -
     \frac{3}{16\pi^2}\beta_h^{(1)}h^3.                \lb{47}
\ee
Integrating with respect to $M$ and using the normalization
condition:
\be
       V^{(2)}(\phi^2 = M^2) = 0,           \lb{48}
\ee
we obtain:
\be
  V^{(2)} = \frac{1}{2}A^{(2)} \phi^4 \log(\frac{\phi^2}{M^2})
    +  \frac{1}{4}B^{(2)} \phi^4 (\log \frac{\phi^2}{M^2})^2.  \lb{49}
\ee
Therefore the two--loop effective potential of the SM for
$$
\phi^2 >> m^2$$
becomes:
\be
  V_{eff}(2-loop) = [\frac{\lambda}{8} + \frac{1}{2}A\log(\frac{\phi^2}{M^2})
    +  \frac{1}{4}B(\log \frac{\phi^2}{M^2})^2]\phi^4 + C,
\lb{50a}
\ee
where $C$ is ``the cosmological constant",
$$B\equiv B^{(2)},$$
and
\be
    A = \frac{1}{8}(\beta_{\lambda}^{(1)} +  \beta_{\lambda}^{(2)}) +
    \frac{\lambda}{2}(\gamma^{(1)} + \gamma^{(2)} + (\gamma^{(1)})^2) +
     \frac{1}{8}\gamma^{(1)}\beta_{\lambda}^{(1)}.
                                    \lb{51a}
\ee
In terms of the evolution variable $t$ (\ref{3}), we have:
\be
 V_{eff}(2-loop) = (\frac{\lambda}{8} + At + Bt^2)\phi^4 + C.
             \lb{50b}
\ee

\section{The second minimum of the effective potential}

In this section our goal is to show the possible existence of a second
(non-standard) minimum of the effective potential in the pure SM at the
fundamental scale:
\be
            \phi_{min2} >> v = \phi_{min1}.      \lb{51b}
\ee
The tree--level Higgs potential with the standard ``weak scale
minimum" at $\phi_{min1} =v$ is given by:
\be
       V(tree-level) = \frac{\lambda}{8}(\phi^2 - v^2)^2 + C.
                                               \lb{52}
\ee
In accord with cosmological results, we take the cosmological
constants $C$ for both vacua equal to zero (or approximately
zero): $C=0$ (or $C\approx 0$). The following requirements must be
satisfied in order that the SM effective potential should have two
degenerate minima:
\be
        V_{eff}(\phi_{min1}^2) = V_{eff}(\phi_{min2}^2) = 0,       \lb{53}
\ee
\be
        V'_{eff}(\phi_{min1}^2) = V'_{eff}(\phi_{min2}^2) = 0,       \lb{54}
\ee where
\be
         V'(\phi^2) = \frac{\partial V}{\partial \phi^2}.
                                             \lb{55}
\ee
These degeneracy conditions first considered in Ref.~\ct{1}
correspond to the MPP expectation. {\it The first minimum is the standard
``Weak scale minimum", and the second one is the non-standard
``Fundamental scale minimum" (if it exists).} An illustrative schematic
picture of $V_{eff}$ is presented in Fig.~1.

Here we consider the SM theory with zero temperature ($T=0$). As
was shown in Ref.~\ct{1}, the above MPP-requirements lead to the
condition that our electroweak vacuum is barely stable at $T=0$,
so that in the pure SM the top quark and Higgs masses should lie
on the SM vacuum stability curve, investigated in Refs.~[27,28].

With good accuracy, the predictions of Ref.~\ct{1} for the top quark
and Higgs masses from the MPP requirement of a second degenerate vacuum,
together with the identification of its position with the Planck scale
$\phi_{min2} = M_{Planck}$, were as follows:
\be
M_t = 173 \pm 5\,\,{\mbox{GeV}},\quad M_H =
135 \pm 9\,\,{\mbox{GeV}}. \lb{56}
\ee
Later, in Ref.~\ct{29}, an alternative metastability requirement for
the electroweak (first) vacuum was considered, which gave a Higgs
mass prediction of $122 \pm 11\,\,$ GeV, close to the LEP lower
bound of 115 GeV (Particle Data Group \ct{30}).

Following Ref.~\ct{1}, let us now investigate the conditions,
Eqs.~(\ref{53},\ref{54}), for the
existence of a second degenerate vacuum at the fundamental scale:
\be
          \phi_{min2}\sim {\mu}_{fundamental}.               \lb{57}
\ee
For large values of the Higgs field,
\be
\phi^2 >> m^2,
\ee
$V_{eff}$ is very well approximated by the quartic term in
Eq.~(\ref{6a}) and the degeneracy condition (\ref{53}) gives:
\be
       \lambda(\phi_{min2}) = 0, \lb{61}
\ee
The condition (\ref{54}) for a turning value then gives:
\be
 \lambda'(\phi_{min2}) = 0    \lb{62}
\ee
which can be expressed in the form:
\be
      \beta_{\lambda}(\phi_{min2}, \lambda=0) = 0.  \lb{63}
\ee

In the next section we search for the scale $\phi_{min2}$ given by
the degeneracy conditions (\ref{61},\ref{63}).

\section{The top quark Yukawa coupling constant evolution}
\lb{top}

The position of the second minimum of $V_{eff}$ essentially depends on the
running of the gauge couplings and of the top quark Yukawa coupling.
Let us consider first the running of the gauge couplings $g',\, g,\,g_3$ in
accord with the present experimental data.

Starting from the Particle Data Group \ct{30}, we have the masses:
\be
      M_t = 174.3 \pm 5.1 \,\, {\mbox{GeV}}, \lb{64}
\ee
\be
      M_Z = 91.1872 \pm 0.0021 \,{\mbox{GeV}}. \lb{65}
\ee
Also for the inverse electromagnetic fine structure constant
in the $\ov{MS}$-scheme we have:
\be
         {\hat \alpha}^{-1}(M_Z) = 127.93 \pm 0.027,
                                    \lb{66}
\ee
while for the square of the sine of the weak angle in the
$\ov{MS}$-scheme we have:
\be
    {\hat s}^2(M_Z) = 0.23117 \pm 0.00016,   \lb{67}
\ee
and for the QCD fine structure constant we have:
\be
       \alpha_3(M_Z) = 0.117 \pm 0.002.
              \lb{68}
\ee The running top quark mass was considered in Refs.~[1,28] and
[31-33], for which we have the following value from
Eq.~(\ref{64}):
\be m_t(M_t) \approx (165\pm 5)\,\,{\mbox{GeV}},
\lb{69}
\ee which is related to the running top quark Yukawa
coupling $h(\mu)$ as follows: \be m_t(M_t)\approx
h(M_t)\frac{v}{\sqrt 2}, \ee that is: \be
            h(M_t)\approx 0.95 \pm 0.03.           \lb{70}
\ee

It is well-known that, for $\mu > M_Z$, the running of all the gauge
coupling constants in the SM is well described by the one-loop
approximation. So, for $\mu > M_t$, we can write:
\be
  \alpha_i^{-1}(\mu) = \alpha_i^{-1}(M_t) + \frac{b_i}{2\pi}
\log\left(\frac{\mu}{M_t}\right),  \lb{71}
\ee
where
$$
       \alpha_Y = \frac{g'^2}{4\pi}, \quad \alpha_2 = \frac{g^2}{4\pi},
\quad \alpha_3 = \frac{g_3^2}{4\pi }
$$
i=Y,2,3 for the $ U(1)_Y$, $SU(2)$ and $SU(3)$ groups, and
\be
    b_Y = - \frac{41}{6},\quad b_2 = \frac{19}{6},\quad b_3 = 7.
                                       \lb{72}
\ee
Assuming that this running is valid up to the ``fundamental" scale,
where new physics enters, we have the following evolutions, which
are revised in comparison with Ref.~\ct{34} using updated
experimental results \ct{30}:
\be
      \alpha_Y^{-1}(t) = 97.40 \pm 0.04 - \frac{41}{12\pi}t,
                                           \lb{73a}
\ee
\be
      \alpha_2^{-1}(t) = 29.95 \pm 0.02 + \frac{19}{12\pi}t,
                                            \lb{73b}
\ee
\be
      \alpha_3^{-1}(t) = 9.336 \pm 0.170 + \frac{7}{2\pi}t.
                                           \lb{73c}
\ee
where
\be
t = \log \left(\frac{\mu}{M_t} \right)
\ee
These gauge coupling constant evolutions are given in Fig.~2, where
$x = \log_{10}\mu \,(GeV)$.

Now we are ready to solve the RGE for the top quark Yukawa
coupling, assuming that it is described by the 1--loop
approximation with good accuracy.
Taking into consideration the Ford-Jones-Stephenson-Einhorn RGE
for h(t) \ct{26}, we obtain the following differential equation
for $y=\alpha_h^{-1}(t) = 4\pi/h^2(t)$ in the 1-loop approximation:
\be
     \frac{dy}{dt} = - \frac{9}{4\pi} + F(t)y(t),     \lb{75}
\ee
where
\be
F(t)= \frac{1}{\pi}\left[\frac{17}{24\alpha_Y^{-1}(t)}
 + \frac{9}{8\alpha_2^{-1}(t)} + \frac{4}{\alpha_3^{-1}(t)}\right].
                              \lb{76}
\ee
Now, using the central values of $\alpha_i^{-1}(M_t)$, we can solve
the RGE for $ y=\alpha_h^{-1}(t)$. We take the spread of
experimental values of $h(M_t)$ in Eq.~(\ref{70}) to give us the
following choice of initial values:
\be
    y(M_t) = \alpha_h^{-1}(M_t) = 13.92,\, 14.85,\, 13.08.
                                                \lb{77a}
\ee
The corresponding solutions for $y(t)$ are presented in Fig.~3 as
the bunches of curves 1, 2 and 3 respectively.
Each bunch describes the spread in the evolution of $y(t)$
due to the uncertainty in $\alpha_3^{-1}(M_t)$ coming from Eq.(\ref{68}):
\be
\alpha_3^{-1}(M_t)=9.336\pm 0.170.
                                                \lb{77b}
\ee
The influence of the uncertainties in $\alpha_{1,2}(M_t)$ is negligible.

The curve y1 of Fig.~3 describes the solution for
$y=\alpha_h^{-1}(t)$ given by the requirement
$\beta_{\lambda}(t,\lambda(t)=0) = 0$ for a second degenerate minimum.
From Eq.~(\ref{29}) in the Appendix, the curve y1 is given by the
equation:
\be
   y_1(t) = \left[\frac{1}{16\alpha_Y^{-2}(t)} +
\frac{1}{8\alpha_Y^{-1}(t)\alpha_2^{-1}(t)}
   +\frac{3}{16\alpha_2^{-2}(t)}\right]^{-\frac{1}{2}},    \lb{80}
\ee in the 1-loop approximation. We note that numerically $y_1(t)$
varies rather weakly as a function of $t$, corresponding to a
value for the top quark Yukawa coupling of \be
h(\mu_{fundamental}) = h(\phi_{min2}) \simeq 0.4          \lb{80b}
\ee at the second or fundamental scale minimum [2,7]. The
intersections of this curve y1 with the possible evolutions of
$y=\alpha_h^{-1}(t)$ in Fig.~3 determine the positions of the
second minimum, according to the MPP assumption. In this way we
obtain the following positions for the second minimum (at
$\mu_0=\phi_{min2}$ and $t_0 = \log(\mu_0/M_t)$), which depend on
the value of $h(M_t)$:

\vspace{0.2cm}

I) $\alpha_3(M_Z) = 0.117;$
$$y(M_t) = 13.92,\quad h(M_t)\approx
0.95,\quad t_0\approx 44.5,\quad  \mu_0\sim 10^{21.5}\,\,
\mbox{GeV}; $$
$$ y(M_t)  = 14.85,\quad h(M_t)\approx 0.92,
\quad t_0\approx 38.8,\quad \mu_0\sim 10^{19}\,\, \mbox{GeV}; $$
$$y(M_t) = 13.08,\quad h(M_t)\approx 0.98,\quad
t_0\approx 50.7,\quad \mu_0\sim 10^{24}\,\, \mbox{GeV};$$

\vspace{0.2cm}

giving the range
$$
             \phi_{min2} \sim 10^{19}-10^{24} \ \mbox{GeV}.
$$

II) $\alpha_3(M_Z) = 0.115;$
$$y(M_t) = 13.92,\quad h(M_t)\approx 0.95,\quad
t_0\approx 42,\quad \mu_0\sim 10^{20.5}\,\, \mbox{GeV}; $$
$$y(M_t) = 14.85,\quad h(M_t)\approx 0.92, \quad t_0
\approx 38.3,\quad \mu_0 \sim 10^{19}\,\, \mbox{GeV}; $$
$$y(M_t) = 13.08,\quad
h(M_t)\approx 0.98,\quad t_0\approx 50,\quad \mu_0\sim 10^{24}\,\,
\mbox{GeV};$$

\vspace{0.2cm}
giving the range
$$\phi_{min2} \sim 10^{19}-10^{24} \ \mbox{GeV}. $$

III) $\alpha_3(M_Z) = 0.119;$
$$y(M_t) = 13.92,\quad h(M_t)\approx 0.95,\quad
t_0\approx 46.3,\quad \mu_0\sim 10^{22}\,\, \mbox{GeV}; $$
$$y(M_t)
= 14.85,\quad h(M_t)\approx 0.92, \quad t_0\approx 40.6,\quad \mu_0\sim
10^{20}\,\, \mbox{GeV}; $$
$$y(M_t) = 13.08,\quad
h(M_t)\approx 0.98,\quad t_0\approx 52.2,\quad \mu_0\sim 10^{25}\,\,
\mbox{GeV};
$$
giving the range
$$\phi_{min2} \sim 10^{20}-10^{25} \ \mbox{GeV}.
$$

We have also calculated the position of the second minimum numerically
in the 2-loop approximation. The positions obtained are given in Fig.~4,
which correspond to the following values:

I) $\alpha_3(M_Z) = 0.117;$
$$y(M_t) = 13.92,\quad h(M_t)\approx
0.95,\quad t_0\approx 38.5,\quad  \mu_0\sim 10^{19}\,\, \mbox{GeV}; $$
$$y(M_t)  = 14.85,\quad
h(M_t)\approx 0.92,\quad t_0\approx 33,\quad \mu_0\sim
10^{17}\,\, \mbox{GeV}; $$
$$y(M_t) = 13.08,\quad h(M_t)\approx 0.98,\quad
t_0\approx 44,\quad \mu_0\sim 10^{21}\,\, \mbox{GeV};$$

\vspace{0.2cm}

giving the range
$$
\phi_{min2} \sim 10^{17}-10^{21} \ \mbox{GeV}.
$$

II) $\alpha_3(M_Z) = 0.115;$
$$
y(M_t) = 13.92,\quad h(M_t)\approx 0.95,\quad t_0\approx
36.5,\quad \mu_0\sim 10^{18}\,\, \mbox{GeV}; $$
$$y(M_t)
= 14.85,\quad h(M_t)\approx 0.92, \quad t_0\approx 32,\quad \mu_0\sim
10^{16}\,\, \mbox{GeV}; $$
$$y(M_t) = 13.08,\quad
h(M_t)\approx 0.98,\quad t_0\approx 42,\quad \mu_0\sim
10^{20}\,\, \mbox{GeV};$$

giving the range
$$
             \phi_{min2} \sim 10^{16}-10^{20} \ \mbox{GeV}.
$$

III) $\alpha_3(M_Z) = 0.119;$
$$
y(M_t) = 13.92,\quad h(M_t)\approx 0.95,\quad t_0\approx
39,\quad \mu_0\sim 10^{19.5}\,\, \mbox{GeV}; $$
$$y(M_t)
= 14.85,\quad h(M_t)\approx 0.92, \quad t_0\approx 34,\quad \mu_0\sim
10^{17}\,\, \mbox{GeV}; $$
$$y(M_t) = 13.08,\quad
h(M_t)\approx 0.98,\quad t_0\approx 45.5,\quad \mu_0\sim
10^{22}\,\, \mbox{GeV};$$

giving the range
$$
             \phi_{min2} \sim 10^{17}-10^{22} \ \mbox{GeV}.
$$

Thus, the curve y1 of Fig.~4 shows that, in the 2-loop approximation,
the experimental values of the coupling constants allow the SM effective
potential to have a second minimum in the interval:
\be
\phi_{min2} \in (10^{16}, 10^{22})\,{\mbox {GeV}}.
\ee

Here we emphasize that, for the central values of the experimental
parameters:
$$
    \alpha_s(M_Z) \approx 0.117 \quad\quad
{\mbox{and}} \quad\quad  h(M_t)\approx 0.95,
$$
and identifying the position of the second minimum with the
fundamental scale, $\mu_{fundamental} = \phi_{min2}$, we predict
the fundamental scale to be close to the Planck scale
$$
               \mu_{fundamental}\sim 10^{19}\,\,{\mbox{GeV}},
$$
which coincides with the result of Refs.~[1,28,31,32]. We note
that for the value $h(M_t)=0.98$, and for the values $h(M_t)=0.95$
and $\alpha_3(M_Z)= 0.119$, the second minimum of the effective
potential turns out to be beyond the Planck scale ($M_{Planck}\sim
10^{19}$ GeV). On the other hand, for the extreme values
$\alpha_s\approx 0.115$ and $h(M_t)\approx 0.92$, the fundamental
scale becomes $\mu_{fundamental}\approx 10^{16}$ GeV,
corresponding to the string scale.

\section{The second derivative of the SM effective potential and
the second minimum at $10^{19}$ GeV }

Let us consider now the second minimum of $V_{eff}$ for the central
values of the experimental parameters, when $\phi_{min2}=10^{19}$ GeV or
$t_0\approx 38.5$.
Choosing the renormalization point at $M = \phi_{min2}$, we
introduce the following evolution parameter:
\be
            t' = \log(\frac{\phi}{\phi_{min2}}).    \lb{85}
\ee
With the definition:
\be
   V \stackrel{def}{=}\frac{(16\pi )^4}{24}(\phi_{min2}^{-4})V_{eff},
                                 \lb{85a}
\ee
we have obtained the following expression for the effective
potential in the two-loop approximation:
\be
V = (C_1 t' + C_2t'^2)(\frac{\phi}{\phi_{min2}})^4 = (C_1 t' + C_2
t'^2)e^{4t'},
                                                  \lb{86}
\ee
where our calculations gave:

\be C_1 = (g'^2 + 3g^2 -4h^2)(\frac{3}{4}g'^4 + \frac{3}{2}g'^2g^2
+ \frac{9}{4}g^4 - 12h^4),
    \lb{87a}
\ee
$$
    C_2 = 2C_1
   + \frac{41}{3}(g'^6 + g'^4g^2) - \frac{19}{3}(3g^6 + g'^2g^4)
$$
\be
   - 32h^4(\frac{9}{2}h^2 - 8g_3^2 - \frac{9}{4}g^2 -
   \frac{17}{2}g'^2).    \lb{87b}
\ee

Calculating all the parameters at $t_0\approx 38.5$, we have:
\be
g'^2(t_0)\approx 0.2263,\quad\quad g^2(t_0)\approx
0.2546,\quad\quad g_3^2(t_0)\approx 0.2406,\quad\quad
  h^2(t_0)\approx 0.1571, \lb{88}
\ee
and
\be
    C_1 \approx -0.00921,\quad \quad
    C_2 \approx 2.8639.                \lb{89}
\ee

\subsection{Hierarchy without use of the new bound state?}

As emphasized in the introduction, our present explanation for the
hierarchy of scales depends crucially on the existence of a third
SM phase associated with the condensation of a proposed
new bound state. However, we first consider here a superficially
appealing argument for the huge scale ratio based just on the
existence of two minima in the effective potential and some
form of ``naturalness''.

Therefore let us consider the second derivative of the effective
potential, which has to change its sign from ``+" to ``-" and
back again to ``+" in the region between the two minima. We take, as
our ``naturalness'' assumption, that in a polynomial approximation
to the second derivative of the effective potential
\be
    V''(t) = \frac{\partial^2 V}{\partial (\phi^2)^2}
= a_0 +a_1 t + a_2 t^2,         \lb{93a}
\ee
where
\be
           t = \log(\phi/M_t) = t' + t_0 \approx t' + 38.5,   \lb{94}
\ee
the coefficients can be considered to be random numbers, but with their
phenomenological orders of magnitude imposed. In the 0-loop
approximation only $a_0$ would be different from zero, in the 1-loop
approximation only $a_0$ and $a_1$ would be non-zero and so on. In fact
the ratios of successive expansion coefficients $r_1 = |a_1/a_0|$,
$r_2 = |a_2/a_1|$,... are expected to be of the order of magnitude
$\beta_{\lambda_p}/\lambda_p$. This expectation expresses the idea
that the variation of $V''$, which in first approximation is
proportional to $\lambda(\phi) = \lambda(t)$, is given by the beta
functions $\beta_{\lambda_p}$ of Eq.~(\ref{24}) measured relative to the
respective couplings $\lambda_p$. Phenomenological values for these
relative rates of variation $\beta_{\lambda_p}/\lambda_p$,
evaluated at the Planck scale say, are typically numbers of the
order of 1/90.

We may check this idea by evaluating the expansion coefficient ratios
$r_1$ and $r_2$ for our expansion Eq.~(\ref{93a}), using
Eqs.~(\ref{86},\ref{89}). Indeed we find the coefficients $a_i$ to be
\be
   a_0 \approx 8161.2,\quad a_1 \approx -432.46,\quad a_2 \approx 5.7277,
                                      \lb{95}
\ee
which give the ratios:
\be
        r_1=|\frac{a_1}{a_0}|\approx 0.0529 \sim
        \frac{1}{19},\quad \quad
        r_2=|\frac{a_2}{a_1}|\approx 0.0132 \sim
        \frac{1}{75}.                   \lb{96}
\ee So we see that these expansion ratios are indeed small, of the
order 1/19 to 1/75; not so very different from the suggested 1/90.

With random coefficients having ratios of this order of magnitude,
we would expect the typical range in $t$ with $V'' < 0$ between
two regions with $V'' > 0$, for a second derivative $V''$ having
such an expansion, should have a length $\Delta t$ of order 19 to
75, or say 90. Consequently there should be a similar range
$\Delta t$ between the first and second minimum of $V$, implying
an exponentially huge scale ratio of the order:
\be
\frac{\phi_{min2}}{\phi_{min1}} \sim \exp(\Delta t) \sim \exp(19 \
\mbox{to} \ 75 \ \mbox{or} \ 90)
\ee
In this way we seem to have
solved the huge scale ratio problem -- here taken to be the ratio
of the Higgs field values in the two assumed minima of the
effective potential.

However we must immediately admit that this solution is not satisfactory!
Imposing the phenomenologically needed small Higgs mass of order
$\phi_{min1}$, the quartic $\phi^4$ term comes to dominate the Higgs
effective potential in most of the region between the two minima and near
$\phi_{min2}$. So then $V''(t)$ roughly functions as the running
self-coupling $\lambda(t)$. But now, generically in the above scenario,
$V''(t)$ is negative in a large interval, which would cause the effective
potential $V_{eff}$ to be negative and make our vacuum severely unstable.

Contrary to this general expectation, our data-based picture with
the degenerate vacuum conditions, Eqs.~(\ref{61},\ref{62}),
imposed is fine-tuned in such a way as to avoid this problem of
negative values for the effective potential. It can indeed be
readily seen that the coefficients $a_i$ of Eq.~(95) are not
random, which would require the distance $\Delta t$ between the
zeros of $V''$ to be of the order of 19 to 75 or 90. However, as
one sees from Fig.~5, in our realistic picture the two zeros of
$V''$ are only about one unit in $t$ apart! This behavior of
$V''(t')$ was determined using the following expansion:
\be
    V''= a t'^2 + b t' + c,         \lb{90}
\ee
where from Eq.~(\ref{86})
\be
     a = 2C_2,\quad b = 2C_1 + 3C_2, \quad
     c = \frac{3C_1 + C_2}{2},                \lb{91}
\ee
and from Eq.~(\ref{89})
\be
   a \approx 5.7277,\quad b \approx 8.5731,\quad c \approx 1.4181.
                                      \lb{91b}
\ee
It is a parabola with the above-mentioned sign changes, having zeros at the
following positions:
\be
      t'_1 = - 1.3074,\quad
      t'_2 = - 0.1894,     \lb{92}
\ee
and
\be
 {\mbox{min}}\, V'' = - 1.7899.
                             \lb{93}
\ee

The shape of the second minimum at $\phi_{min2}=10^{19}$ GeV is
presented in Fig.~6 for V given by Eq.~(\ref{85a}).

Near the minimum of V at $10^{19}$ GeV, the Higgs self-coupling
$\lambda(t')$ is well described by the expression:
\be
     \lambda(t')G^4(t') = \frac{192}{(16\pi)^4}(C_1 t' + C_2 t'^2).
                                              \lb{97}
\ee The behavior of $\lambda(t')$ near the second minimum is shown
in Fig.~7 for different values of $h(M_t)$. The behavior of
$\lambda(t)$ at lower energies was obtained and shown in
Refs.~[1,35].

In concluding this part of our paper, we should
emphasize that the MPP-description of the SM predicts a value for the
Higgs mass. In spite of the large uncertainty in the position of the
second minimum, the value of $\lambda$ at the electroweak scale is
predicted to lie in the narrow interval:
\be
            \lambda(M_t)\in (0.27;\,0.34), \lb{98}
\ee
which corresponds to the prediction of the Higgs mass given in
Ref.~\ct{1}: $M_H\approx 135\pm 9$ GeV.
In this scenario new physics enters at a scale of the order
of $10^{19}$ GeV.

\section{A new bound state in the SM}

The MPP helps to solve fine-tuning problems; in particular
the hierarchy problem of why the electroweak scale is so tiny in comparison
with the Planck scale.

It is well-known that, when calculating the square of the SM Higgs
mass, we have to deal with the quadratic divergencies which occur
order by order in the perturbative expansion. The bare Higgs mass
squared needs to be fine-tuned in all orders of this perturbation
series. Near the (cut-off) Planck scale these quadratic
divergencies become $(\Lambda_{Planck}/\Lambda_{weak})^2$ (that
is, $(10^{17})^2 =10^{34}$) times bigger than the final Higgs mass
squared, and it is clear that a fine-tuning by 34 digits is needed
to solve the hierarchy problem in the SM. Supersymmetry can remove
these divergencies by having a cancellation between fermion and
boson contributions. Hence supersymmetry solves the technical
hierarchy problem. But the problem still exists in the form of why
the soft supersymmetry breaking terms are small compared to the
fundamental Planck scale.

At first sight, it seems difficult to explain the cancellation of
such divergencies by the MPP protected fine-tuning. The energy
density, or cosmological constant, has the dimension of energy to
the fourth power, so that modes with Planck scale frequencies
contribute $(10^{17})^4=10^{68}$ times more than those at the
electroweak scale. Nevertheless, we can obtain such a fine-tuning
by assuming the existence of two degenerate phases in the SM which
are identical for the highest modes, but deviate by their physics
at the electroweak scale. Thus the way to solve the hierarchy
problem in the SM using MPP is to consider a new phase at the
electroweak scale, different from and degenerate with the
Weinberg-Salam Higgs phase. The obvious way to achieve this is to
find  a scalar bound state $\phi_{bound}$ made out of SM
particles, which is so strongly bound that it becomes tachyonic
and condenses into the vacuum. Of course we do not expect that the
tachyons should really exist in Nature, but that the vacuum
condensate should adjust itself to such a density as to bring the
mass squared of the bound state back to be positive. As was shown
in Refs.~[2,3,7], an attractive candidate for such a bound state
could be composed from $6t + 6\bar t$ quarks.

Here the Weinberg-Salam Higgs particle exchange plays an essential
part. The virtual exchange of Higgs scalar bosons between $qq$,
$q\bar q$ and $\bar q \bar q$ yields an attractive force in all
cases. The bound state of a top quark and an anti-top quark
(toponium) is mainly bound by gluon exchange, although Higgs
exchange is comparable. But if we add more top or anti-top
quarks, then the Higgs exchange continues to attract while the
gluon exchange saturates and gets less significant. The maximal
binding energy per particle comes from the S-wave $6t+6\bar t$
ground state. The reason is that the $t$ quark has 2 spin states
and 3 colour states. This means that, by the Pauli principle, only
$2\times 3=6 \ t$ quarks can be put in an S-wave function,
together with $6 \ \bar t$-quarks. So, in total, we have 12 $t$
quark/anti-quark constituents together in relative S-waves. If we
try to put more $t$ and $\bar t$ quarks together, then some of
them will go into a P-wave for which the pair binding energy will
decrease by at least a factor of 4.

Estimating the pair binding energy using the Bohr formula for
atomic energy levels (here for each quark, we treat the remaining
11 quarks as a nucleus), the authors of Refs.~[2,7] have obtained
an approximate expression for the mass squared of the $6t+6\bar t$
bound state. Their result for the mass squared of the bound state,
crudely estimated from the non-relativistic binding energy, is:
\be
   m_{bound}^2 = (12m_t)^2 - 2(12m_t) \times E_{binding} +...
   \approx (12m_t)^2 ( 1 - \frac{33}{8\pi^2}h^4
        +...),                       \lb{99}
\ee
The condition that this bound state should be tachyonic leads
to the requirement:
\be
         (1 - \frac{33}{8\pi^2 }h^4 + ....) \le 0.
                                                  \lb{100}
\ee
When the bound state becomes tachyonic, we should be
in a vacuum state with
\be
              <\phi_{bound}>\neq 0.    \lb{101}
\ee
Hence we expect a phase transition to the new phase at the point where
the bound state mass squared passes zero as a function of the top quark
Yukawa coupling $h$:
\be
                    m^2_{bound} = 0,      \lb{102}
\ee According to Refs.~[2,3,7], this condition gives:
\be
     h_{p.t.}\approx \sqrt{\pi \sqrt{\frac{8}{33}}}\approx 1.24,
                                                 \lb{103}
\ee
where ``p.t." means the value at the ``phase transition".

Taking into account a possible correction due to the Higgs field
quantum fluctuations, we gave the following estimate in Ref.~\ct{3}:
\be
           h_{p.t.}(\mu_{electroweak}) = 1.06\pm 0.18,            \lb{104}
\ee
which is in agreement with the experimental value of the top quark
Yukawa coupling constant at the electroweak scale: $h_{exper}(M_t)
= 0.95 \pm 0.03$. It seems that we have a successful
confirmation of the MPP hypothesis based on pure SM physics
at the electroweak scale. In principle a more accurate (but hard)
SM calculation of the top quark Yukawa coupling at the phase
transition, $h_{p.t.}$, would provide a very clean test of MPP.

Now we have {\it not two, but three vacua in the SM with the same energy density}
and in the next section we explain how they lead to a resolution
of the hierarchy problem in the SM.

\section{The hierarchy of scales in the SM}

Requiring the same energy density of three SM vacua (new,
standard and fundamental), we have obtained predictions for the
running top quark Yukawa coupling $h(t)$ at the fundamental scale
$\mu_{fundamental}$, Eq.~(\ref{80b}), and the electroweak
scale $\mu_{electroweak}$, Eq.~(\ref{104}). So we can now use
the RGE for the top quark Yukawa coupling to estimate the ratio
of the fundamental and electroweak scales needed to generate
the required amount of renormalization group running of $h(t)$.
It is assumed here that we can take the values of the SM
fine structure constants at the fundamental scale,
$\alpha_i(\mu_{fundamental})$, as given and, in particular, we take
$\alpha_3(\mu_{fundamental}) \simeq 1/54$ (see Fig.~2). Due to the
relative smallness of the beta function $\beta_h(\mu_{fundamental})$
at the fundamental scale, we need many $e$-foldings between
$\mu_{fundamental}$ and $\mu_{electroweak}$. For definiteness, let us
assume that the MPP prediction, Eq.~(\ref{104}), is correct and
coincides with the experimental value $h(\mu_{electroweak}) \simeq 0.95$.
Then we can use the results of section 4 to give the MPP
prediction for the ratio between the fundamental and electroweak scales:
\be
    \frac{\mu_{fundamental}}{\mu_{electroweak}}\sim 10^{17}
\sim e^{40}.
                                                              \lb{105}
\ee
This exponentially huge scale ratio provides our MPP solution to the
hierarchy problem in the SM. In the scenario developed in this
paper we essentially have  a ``great desert'' between the electroweak and
fundamental scales: no new physics, with the exception perhaps of
neutrinos at the see-saw scale.

\section{Conclusions}

As a way of predicting the ratio of the fundamental (Planck) scale to
the electroweak scale, we have developed the idea of the Multiple
Point Principle which states that there exist several different
vacuum states in Nature having the same energy density; more
precisely, all vacua having approximately zero energy density or
cosmological constant.

Neglecting mass terms, we have used the RGE for the SM effective
potential and obtained an explicit expression for this effective
potential in the two-loop approximation for large values of the
Weinberg-Salam Higgs scalar field: $\phi^2 >> m^2$.

We have considered the running of the gauge and top quark Yukawa
couplings, investigating the conditions for the existence of a
second minimum of the effective potential in the pure SM. Our
investigation of the evolution of the top quark Yukawa coupling constant
showed that the experimentally established value at the electroweak
scale, $h(M_t)=0.95\pm 0.03$, is consistent with a second minimum
of the SM effective potential existing in the interval: $$
\phi_{min2}\in (10^{16}, 10^{22})\ {\mbox{GeV}.}$$ The central
experimental values $h(M_t)=0.95$ and $\alpha_3(M_Z) = 0.117$,
together with the vacuum degeneracy conditions (\ref{61},\ref{63}),
predict a second minimum at $\phi_{min2}\approx 10^{19}$ GeV.
We presented the shape of this second minimum, showing the sign changes
of the second derivative of the SM effective potential in the region
$\mu \le \phi_{min2}$.

In the framework of the MPP solution of the hierarchy problem, we
have investigated the possible existence of a very
strongly bound state of 6 top quarks and 6 anti-top quarks due, in the
main, to Weinberg-Salam Higgs boson exchanges between pairs of
quarks/anti-quarks. This $6t + 6\bar t$ bound state is
supposed to condense in a new phase of the SM vacuum, for which
$<\phi_{bound}>\sim \mu_{electroweak}$. An estimate of the top quark
Yukawa coupling $h_{p.t.}$ at the critical point of the phase
transition between the ``new'' phase and the ``weak'' phase revealed
that MPP may explain why $h(M_t) \approx 1$, in accord with
experiment.

We have shown that the requirement of the degeneracy of the three
vacua (new, weak and fundamental) leads to the prediction
of an exponentially huge scale ratio:
$$
\frac{\mu_{fundamental}}{\mu_{electroweak}}\sim e^{40},
$$
in the absence of new physics between the electroweak
and fundamental scales (with the exception of neutrinos).

\section{Acknowledgements}

This work was supported by the Russian Foundation for Basic
Research (RFBR), project No.02-02-17379. L.V.L. and H.B.N. thank
R.Barbieri and the directors of the Conference on Hierarchy
Problems in Four and More Dimensions (Italy, Trieste, 1-4 October,
2003) for the wonderful organization of the Conference and useful
discussion of the talk based on this investigation. L.V.L. thanks
all participants at her Niels Bohr Institute theoretical seminar,
especially P.H.Damgaard, D.Diakonov, N.Obers and F.Sannino, for
fruitful discussions and comments. We also thank our colleagues
D.L.Bennett and R.B.Nevzorov for their helpful discussions and
advice.

\section{Appendix}

In this Appendix we give the results of Ref.~\ct{26} for the RGE
$\beta$-functions in the one-loop and two-loop approximations.

The one-loop beta-functions in the SM are:
\be
16\pi^2 \beta_{g'}^{(1)} = (\frac{10}{9}N_f + \frac{1}{6}N_S)
g'^3,                     \lb{25}
\ee
\be
16\pi^2 \beta_{g}^{(1)}
= (\frac{2}{3}N_f + \frac{1}{6}N_S - \frac{22}{3}) g^3, \lb{26}
\ee
\be
16\pi^2 \beta_{g_3}^{(1)} = (\frac{2}{3}N_f - 11) g_3^3,
\lb{27}
\ee
and
\be 16\pi^2 \beta_{h}^{(1)} = h(\frac{9}{2} h^2 -
8g_3^2 - \frac{9}{4} g^2 - \frac{17}{12} g'^2), \lb{28}
\ee
\be
16\pi^2\beta_{\lambda}^{(1)} = 12\lambda^2 + \lambda (12h^2 - 9g^2
- 3g'^2) + \frac{3}{4}g'^4 + \frac{3}{2}g'^2 g^2 + \frac{9}{4}g^4
- 12h^4, \lb{29}
\ee
\be
\beta_{m^2}^{(1)} = m^2
(2\gamma^{(1)} + \frac{3\lambda}{8\pi^2}),         \lb{30}
\ee
where
\be
64\pi^2 \gamma^{(1)} = 3(g'^2 + 3g^2 - 4h^2)
                              \lb{31}
\ee is the one--loop anomalous dimension in the Landau gauge. Here
we use $N_f$ for the number of flavors, and $N_S$ for the number
of the Higgs doublets.

In the region $\mu > M_t$, where $M_t$ is the top quark pole mass,
we have:
\be
      N_f=6,\quad N_S=1,        \lb{32}
\ee
resulting in the following RGE beta-functions for the gauge
coupling constants in the one-loop approximation:
\be
16\pi^2
\beta_{g'}^{(1)} = \frac{41}{6}g'^3, \lb{33} \ee \be 16\pi^2
\beta_{g}^{(1)} = - \frac{19}{6}g^3, \lb{34} \ee \be 16\pi^2
\beta_{g_3}^{(1)} = - 7 g_3^3. \lb{35}
\ee

The two--loop contributions to the RG $\beta$-functions are given
by: \be (16\pi^2)^2 \beta_{g'}^{(2)} = g'^3(\frac{199}{18}g'^2 +
\frac{9}{2}g^2 + \frac{44}{3}g_3^2 - \frac{17}{6}h^2), \lb{36} \ee
\be (16\pi^2)^2 \beta_{g}^{(2)} = g^3(\frac{3}{2}g'^2 +
\frac{35}{6}g^2 + 12g_3^2 - \frac{3}{2}h^2), \lb{37} \ee \be
(16\pi^2)^2 \beta_{g_3}^{(2)} = g_3^3(\frac{11}{6}g'^2 +
\frac{9}{2}g^2 - 26g_3^2 - 2h^2),
 \lb{38}
\ee
$$
(16\pi^2)^2 \beta_{h}^{(2)} = h(-12h^4 + \frac{3}{2}\lambda^2 -
6\lambda h^2 + (\frac{131}{16}g'^2 + \frac{225}{16}g^2 + 36
g_3^2)h^2 + $$
\be
 \frac{1187}{216}g'^4 - \frac{3}{4}g'^2g^2 +
\frac{19}{9}g'^2g_3^2 - \frac{23}{4}g^4 + 9g^2g_3^2 - 108g_3^4),
        \lb{39}
\ee
$$
(16\pi^2)^2 \beta_{\lambda}^{(2)} = - 78\lambda^3 +
18\lambda^2(g'^2 + 3g^2 - 4h^2) + \lambda[10(\frac{17}{12}g'^2 +
\frac{9}{4}g^2 + 8g_3^2)h^2 -
$$
$$
\frac{73}{8}g^4 + \frac{39}{4}g'^2g^2 + \frac{629}{24}g'^4 - 3h^4]
+ \frac{305}{8}g^6 - \frac{289}{24}g'^2g^4 - \frac{559}{24}g'^4g^2
- \frac{379}{24}g'^6 -
$$
\be
16(4g_3^2 + \frac{1}{3}g'^2)h^4 + (-\frac{19}{2}g'^4 + 21 g'^2g^2
- \frac{9}{2}g^4)h^2 + 60 h^6,
                  \lb{40}
\ee
and
$$
 -(16\pi^2)^2 \gamma^{(2)} = \frac{3}{2}\lambda^2 -
 \frac{27}{4}h^4 + 20g_3^2h^2 + \frac{45}{8}g^2h^2 +
 \frac{85}{24}g'^2h^2 -
$$
\be
\frac{271}{32}g^4 + \frac{9}{16}g^2g'^2
 + \frac{431}{96}g'^4.
                  \lb{41}
\ee

\newpage

\begin{figure}[t] \centerline{\epsfxsize=\textwidth
\epsfbox{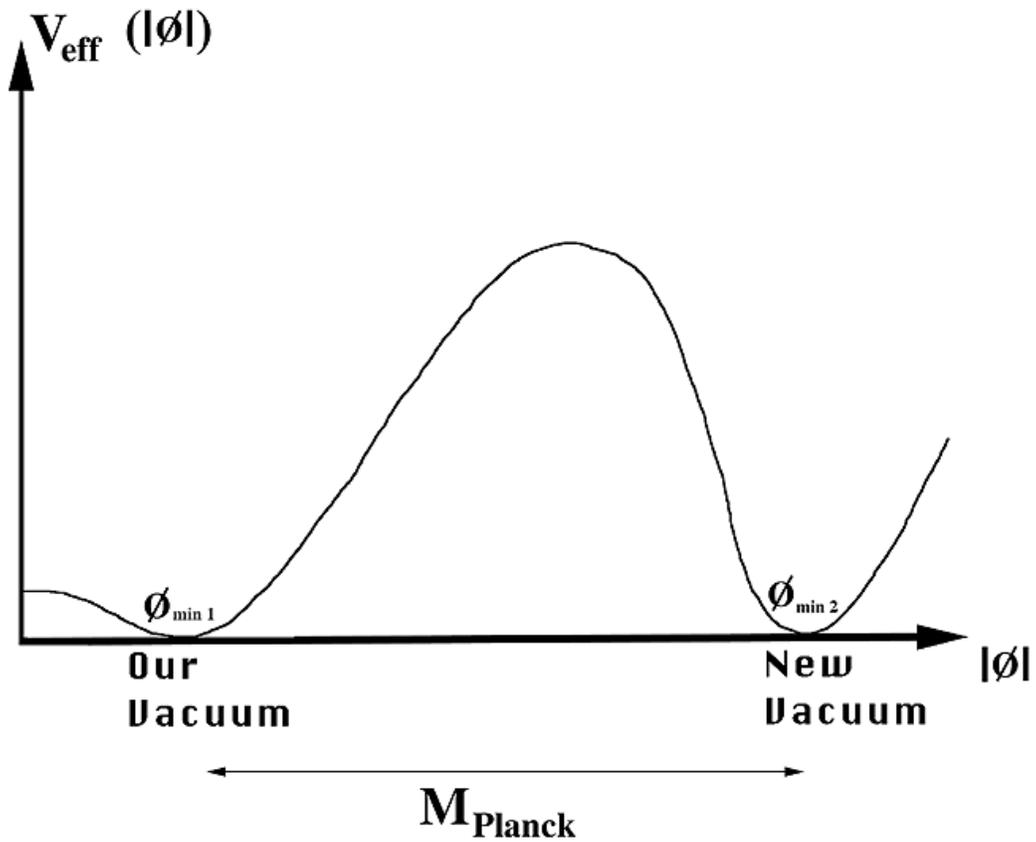}} \caption{\Large The first (our) vacuum at
$|\phi|\approx 246$ GeV and the second vacuum at the fundamental
scale $|\phi|\sim M_{Pl}.$}
\end{figure}

\begin{figure}[t] \centerline{\epsfxsize=\textwidth
\epsfbox{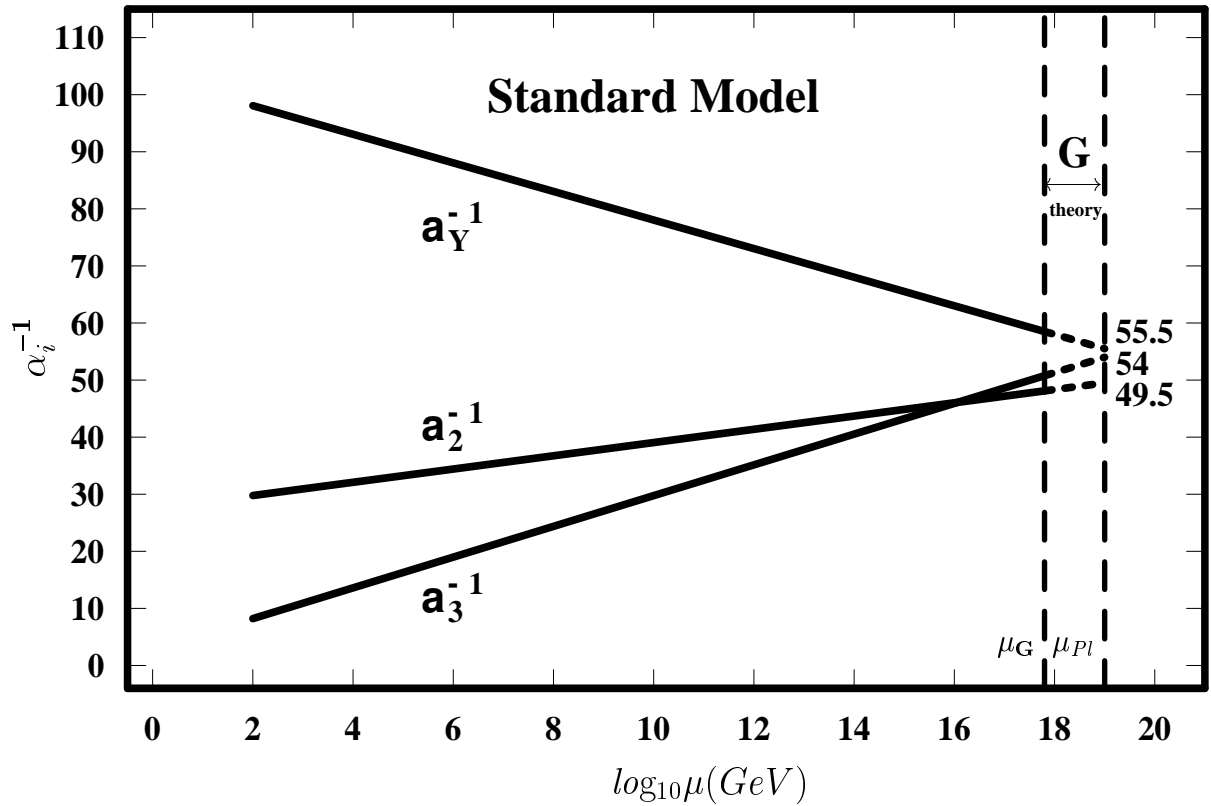}} \caption{\Large The evolution of the
inverse SM fine structure constants
$\mbox{a}_{Y,2,3}^{-1} \equiv \alpha_{Y,2,3}^{-1}$ as functions
of x ($\mu=10^x$ GeV) up to the scale $\mu_G\sim M_{Pl}$ where
new physics (``G-theory") enters.}
\end{figure}

\begin{figure}[t]
\centerline{\epsfxsize=\textwidth \epsfbox{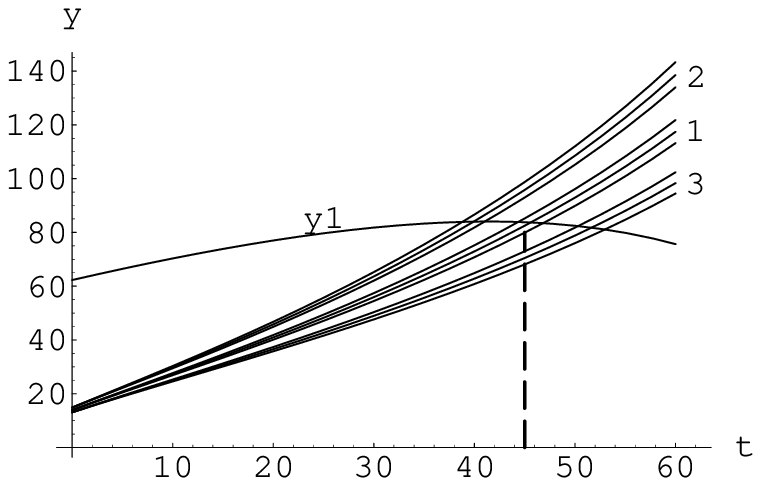}}
\caption{{\Large\bf The 1--loop approximation} for the evolution
of $ y(t)=\alpha_h^{-1}(t)=4\pi/h^2(t)$, where h is the top quark
Yukawa coupling constant. Three bunches 1(middle), 2(up), 3(down)
of curves correspond respectively to the three values of
$h(M_t)=0.95,\,0.92,\,0.98$ given by experiment. The spread of
each bunch corresponds to the experimental values of
$\alpha_3(M_Z) =0.117\pm 0.002$ (upper and lower curves correspond
to $\alpha_3(M_Z)=0.115$ and $\alpha_3(M_Z)=0.119$ respectively).
The curve y1 for $ y_1$, given by the MPP requirement
$\beta_{\lambda}(\lambda=0)=0$, intersects the bunches at points
corresponding to the position of the second minimum identified
with the fundamental scale in the SM.}
\end{figure}

\begin{figure}[t]
\centerline{\epsfxsize=\textwidth \epsfbox{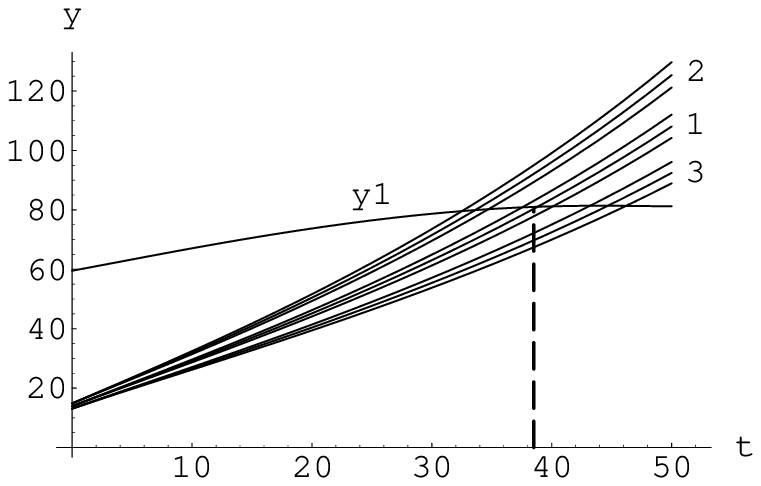}}
\caption{{\Large \bf The 2--loop approximation} for the evolution
of $ y(t)=\alpha_h^{-1}(t)=4\pi/h^2(t)$, where h is the top quark
Yukawa coupling constant. Three bunches 1(middle), 2(up), 3(down)
of curves correspond respectively to the three values of
$h(M_t)=0.95,\,0.92,\,0.98$ given by experiment. The spread of
each bunch corresponds to the experimental values of
$\alpha_3(M_Z)=0.117\pm 0.002$ (upper and lower curves correspond
to $\alpha_3(M_Z)=0.115$ and $\alpha_3(M_Z)=0.119$ respectively).
The curve y1 for $ y_1$, given by the MPP requirement
$\beta_{\lambda}(\lambda=0)=0$, intersects the bunches at points
corresponding to the position of the second minimum identified
with the fundamental scale in the SM.}
\end{figure}

\begin{figure}[t]
\centerline{\epsfxsize=\textwidth \epsfbox{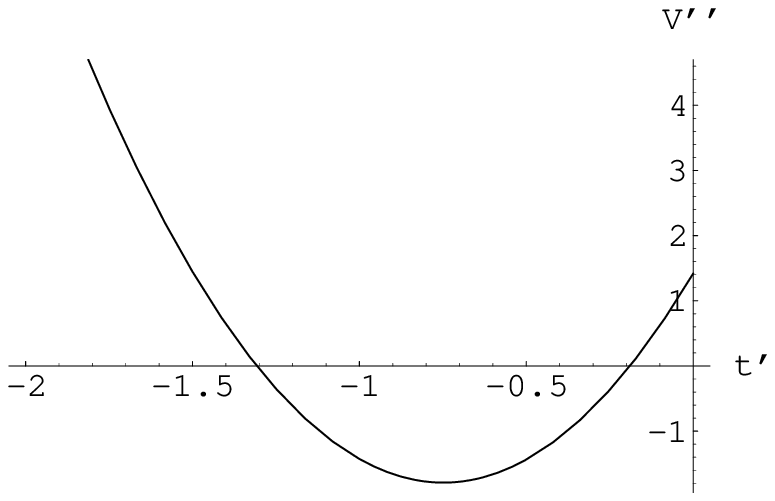}} \caption{
 The behaviour of the second derivative V" ($V\stackrel{def}{=}
\frac{(16\pi)^4}{24}(\phi_{min2}^{-4})V_{eff} $) near the position
of the second minimum of the effective potential at
$\phi_{min2}=10^{19}$ GeV; $t'= t-38.5$.}
\end{figure}

\newpage

\begin{figure}[t]
\centerline{\epsfxsize=\textwidth \epsfbox{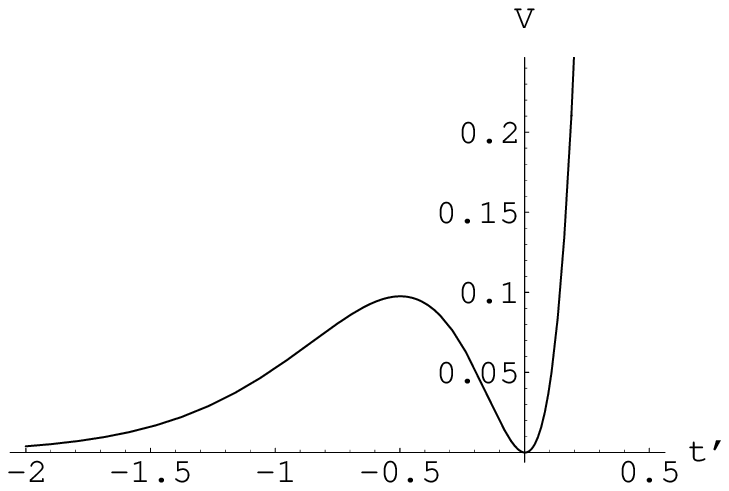}} \caption{
The behaviour of $V \stackrel{def}{=}
\frac{(16\pi)^4}{24}(\phi_{min2}^{-4})V_{eff}$ near the position
of the second minimum of the effective potential at $\Large
\phi_{min2}=10^{19}$ GeV; $t'=t-38.5$.}
\end{figure}

\begin{figure}[t]
\centerline{\epsfxsize=\textwidth \epsfbox{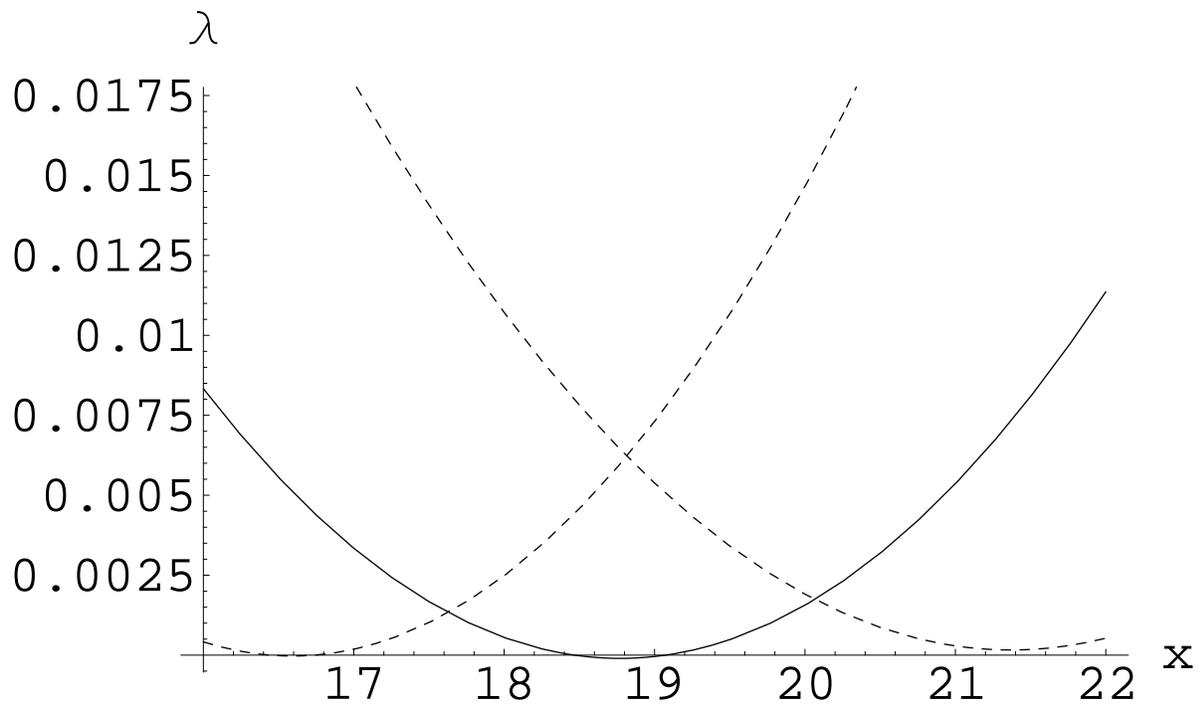}}
\caption{The behaviour of the Higgs self-coupling $\lambda(t')$ near
the second minimum (at $x = \log_{10} \mu \approx 19\pm 2$) corresponding
to the experimental values $h(M_t)=0.95\pm 0.03.$}
\end{figure}

\end{document}